\documentclass[pra,twocolumn,showpacs,showpacs]{revtex4}
\usepackage{graphicx,amsmath,amssymb,amsfonts}

\def\bk{\mathbf{k}}

\begin{document}

\title{Angular momentum of focused beams: beyond the paraxial approximation}
\author{Paula B. Monteiro, Paulo  A. Maia Neto and  H. Moys\'es Nussenzveig}
\address{Instituto de F\'{\i}sica, UFRJ, 
CP 68528,   Rio de Janeiro,  RJ, 21941-972, Brazil}

\begin{abstract}
We investigate in detail the focusing of a circularly polarized Laguerre-Gaussian laser beam ($\hbar \ell$ orbital angular momentum per photon; $\sigma=1/-1$ for left/right-handed polarization) by a high numerical aperture objective. The diffraction-limited focused beam has unexpected properties, resulting from a strong interplay between the angular spatial structure and the local polarization in the non-paraxial regime. In the region near the beam axis, and provided that $|\ell|\ge 2$ and $\ell$ and $\sigma$ have opposite signs, the  energy locally counter-propagates  and the projection of the electric field onto the focal plane counter-rotates with respect to the circular polarization of the incident beam. We explicitly show that the total angular momentum flux per unit power is conserved after focusing, as expected by rotational symmetry, but the spin and orbital separate contributions change. 
\end{abstract}


\date{\today}

\maketitle

\section{Introduction}

Strongly focused laser beams, produced by passage through a high numerical aperture (NA) microscope objective, have been employed increasingly often in recent years. A widespread application is their use for trapping micro-particles \cite{Ashkin86} or atoms \cite{Chu86} in optical tweezers, employed in a variety of fields \cite{Ashkin_livro}, ranging from cell biology \cite{review_biology_applications} to quantum information processing \cite{Grangier}. 

In the standard optical tweezers setup, the objective entrance port is usually illuminated by a Gaussian 
linearly polarized beam, overfilling the entrance aperture in order to fully utilize the high NA of the objective. A proper description of such a tightly focused laser beam must include diffraction at the aperture edge as well as nonparaxial effects. The vectorial Debye-type integral representation developed by Richards and Wolf (RW)  \cite{RWprelim, RW} takes both effects into account. This model has been thoroughly verified directly, by measuring the electric energy density distribution in the focal region \cite{Dorn2003}, and also indirectly, by testing the Mie-Debye theory for the trapping force in optical tweezers (derived from the RW model \cite{tweezers_th}) against experimental results \cite{tweezers_exp}. 

Novel features, with potential applications in micromechanics, microfluidics and biotechnology, have 
been developed by employing Laguerre-Gaussian (LG) beams \cite{Grier_Nature}. These beams have helical wave fronts and a phase singularity along the axis (optical vortex) \cite{Baranova83}. Since paraxial LG beams have enhanced gradient forces and reduced radiation pressure, they produce more efficient trapping \cite{Ashkin}. More importantly, since they carry orbital angular momentum \cite{Allen92} (see also \cite{livro, Bekshaev} for reviews), they allow rotational control of the trapped particles. By transferring orbital angular momentum to absorptive particles, such particles were set to rotate \cite{Friese96}. The photon spin, which is associated with the beam circular polarization, can be employed to add or subtract from the effect of the orbital angular momentum \cite{Friese_and_Simpson}. In some cases, the particle can be trapped off-axis, near the ring of maximum intensity, and set to rotate about the beam axis \cite{ONeil}. By using a spatial light modulator, it is possible to scan the orbital index $\ell$ up to values of order $200$ and analyze the resulting rotation about the beam symmetry axis as a function of $\ell$ \cite{GrierPRL2003}.

We consider a circularly polarized (paraxial) Laguerre-Gaussian  $\rm LG_{0\ell}$ (radial index $p=0$) model for the beam incident along the positive $z$-axis {\it  before} the objective, and then apply the RW approach to calculate the resulting strongly focused beam beyond the objective. Ganic, Gan and Gu calculated the electric energy density on the focal plane for a linearly-polarized LG beam \cite{Gu2003} and showed that the intensity at the focus does not vanish when $\ell=1$ or $\ell=2$. This effect becomes stronger for a circularly polarized incident beam \cite{Bokor2005, Iketaki2007}, with the photon spin anti-parallel to  the orbital angular momentum (the intensity at the focus vanishes when they are parallel). The experimental verification of this remarkable effect provided an additional test of the RW approach  \cite{Bokor2005,Iketaki2007}.

Here we present a detailed theoretical analysis of the focused beam, elucidating the spatial variations of field polarization, energy density and Poynting vector. We find some unexpected results when the spin and orbital angular momenta of the incident beam are antiparallel. In this situation, the interplay between 
polarization and angular dependence is particularly interesting, resulting in a strong modification of the focused beam polarization near the beam axis. 

 The discussion of angular momentum is of particular theoretical relevance. The identification of separate spin and orbital contributions to the angular momentum of electromagnetic fields is somewhat controversial \cite{Cohen}, at least in the sense that the corresponding operators do not satisfy the commutation relations of `true' angular momenta in the full quantum theory \cite{VanEnk}. Nevertheless, 
the separation presents itself in a very compelling way in the paraxial approximation, with spin and orbital contributions corresponding to the field polarization and the angular spatial dependence, respectively \cite{Allen92, Berry}. With $\sigma = +1(-1)$ corresponding to left(right)-handed circular polarization, the spin and orbital angular momenta along the propagation direction per photon are $\sigma\hbar$ and $\ell\hbar$, respectively. In the framework of the classical (i.e. non-quantal) paraxial theory, these important results are inferred from the values of $\sigma/\omega$ and $\ell/\omega$ for the ratios between the linear densities of angular momenta and energy. 

In the nonparaxial regime the separation is not so straightforward, as far as local quantities  are concerned \cite{Barnett94}. However, a natural separation was proposed by Barnett \cite{Barnett02} in terms of the overall angular momentum flux along the propagation direction. 

We calculate the angular momentum flux for the focused nonparaxial beam, and compare the separate spin and orbital  contributions, as defined by Barnett, with the corresponding values for the paraxial beam before the objective. Recently, Zhao and co-workers have argued that the focusing effect leads
to an interconversion between spin and orbital angular momenta \cite{Zhao2007}. We derive quantitative results for the spin and orbital flux modifications, which turn out, however, to have signs in disagreement with Zhao {\it et al}'s qualitative discussion. Since their arguments, as well as their experimental demonstration, involve local quantities, we attribute the disagreement to the fact that we calculate the {\it global} overall flux across a plane perpendicular to the propagation direction, rather than local results. 
  
The paper is organized as follows. We analyze the local polarization on the focal plane in Sec. II, and discuss the energy and energy flux densities in Sec. III. The optical angular momentum is analyzed in Sec. IV, and concluding remarks are presented in Sec. V.
\bigskip

\section{Electric and magnetic fields}

We assume that the beam waist (radius $=w_0$) is precisely at the position of the objective entrance port. The incident circularly polarized electric field  is then given by (the factor $e^{-i\omega t}$ is omitted)
\begin{equation} \label{port}
{\bf E}_{\rm inc}(\rho,\phi,z)= E_{\rm inc}\left( \frac{\sqrt{2}\rho}{w_0} \right)^{|\ell|}\,e^{-\rho^2/w_0^2}
e^{i(\ell\phi+k_0z)}(\mathbf{\hat x}+i\sigma \mathbf{\hat y})
\end{equation}

The minimum spot size $w_0$ is typically of the order of or larger than the entrance aperture radius (overfilling), so as to take full advantage of the high NA of the objective. The focused beam (focal length $f$) is written as a superposition of plane waves. According to Kirchhoff's approximation of classical diffraction theory \cite{RWprelim}, the amplitude and phase of each plane wave component $\mathbf{k}(\theta,\varphi)$ ($|\mathbf{k}|= n k_0 =$ wavenumber in the isotropic medium of refractive index $n$)
are determined by the field at the entrance port ${\bf E}_{\rm inc}(\rho,\phi,z)$ \cite{RWprelim}, with: 
\begin{eqnarray}
\nonumber
\mathbf{E}(\mathbf{r}) &
= & (-1)^{\ell+1}\, \frac{i  E_{0}}{\pi}
\int_{0}^{2\pi}d\varphi e^{i\ell\varphi} \int_{0}^{\theta_{0}}d{\theta}\sin \theta \sqrt{\cos \theta}
\\
 \label{focusedfield}
 & & 
\times (\sqrt{2}\gamma \sin\theta)^{|\ell|}  e^{-\gamma^2\sin^2\theta}
e^{i\mathbf{k}\cdot\mathbf{r}}\,\boldsymbol{\hat{\epsilon}'}_{\sigma}(\theta,\varphi)\\
\label{E0}
E_0 & = & \frac{\pi  f}{\lambda}\,T_{\rm obj}\,E_{\rm inc}, \;\; \gamma = f/w_0, \;\; 
\boldsymbol{\hat{\epsilon}'}_{\sigma}= \mathbf{\hat x}'+i\sigma \mathbf{\hat y}'
\end{eqnarray}
where $\rho = f \sin\theta$ (Abbe sine condition)  and $\phi=\varphi+\pi$. We have assumed that the objective transmission amplitude $T_{\rm obj}$ is uniform. Radial dependence of objective transmittance, which may be present, can be taken into account by introducing an effective beam waist size \cite{transmittance}. 

If the beam is Gaussian at the entrance port ($\ell=0$), all plane wave components in (\ref{focusedfield}) have
 the same phase at the focal position $\mathbf{r} =0$ and hence interfere constructively to produce a maximum intensity at this point. On the other hand, for a Laguerre-Gaussian beam, each component contains the additional phase factor $e^{i\ell\varphi}$, and then the intensity at the focus vanishes except for some special cases discussed below (see also Refs. \cite{Gu2003, Bokor2005, Iketaki2007}).  

In order to compute the field from (\ref{focusedfield}), we need the unit vectors $\mathbf{\hat x}'$ and $\mathbf{\hat y}'$. They are defined in the plane perpendicular to $\bk$ by the condition that the angles between $\mathbf{\hat x}'$ and $\mathbf{\hat y}'$ and the meridional plane at the angular position $\varphi$ are $\varphi$ and $\pi/2-\varphi,$ respectively \cite{foot_unitvectors}. After integration over $\varphi$, we find 
\begin{eqnarray}
E_{x}(\rho,\phi,z) &=& (-i)^{\ell+1} E_0  \left( e^{i\ell\phi}  I^{(\ell)}_{0} +  e^{i(\ell+2\sigma)\phi} I^{(\ell)}_{2\sigma}  \right) \label{Ex} \\
E_{y}(\rho,\phi,z) &=&  \sigma (-i)^{\ell} E_0 \left(e^{i\ell\phi}  I^{(\ell)}_{0} - e^{i(\ell +2\sigma)\phi}  I^{(\ell)}_{2\sigma}  \right) \label{Ey} \\
E_{z}(\rho,\phi,z)  &=&  - 2\sigma (-i)^{\ell} E_0  e^{i (\ell + \sigma)\phi}  I^{(\ell)}_{\sigma}  .
\label{Ez}
\end{eqnarray}
The dependence on $\rho$ and $z$ is contained in the coefficients $I^{(\ell)}_m,$ $m=0,\pm 1,\pm 2,$
which can be expressed in terms of the cylindrical Bessel functions $J_n(\xi)$ \cite{Abramowicz}:
\begin{eqnarray}\label{I}
I^{(\ell)}_{m}(\rho,z) & = & (\sqrt{2} \gamma)^{|\ell|}
  \int_{0}^{\theta_0}  d\theta \, (\sin{\theta})^{|\ell|+1}
\sqrt{\cos\theta}  \\
& & \times e^{-\gamma^2\sin^2\theta}
f_{|m|}(\theta) 
 J_{\ell+m} (k \rho \sin{\theta}) e^{i k z \cos\theta} \nonumber 
\end{eqnarray}
The angular functions $f_m(\theta)$ are given by
\begin{eqnarray}
f_0(\theta) &=& 2 \cos^2(\theta/2) \label{f0} \\
f_1(\theta) &=& \sin\theta  \label{f1} \\
f_2(\theta) &=& 2 \sin^2(\theta/2).  \label{f2}
\end{eqnarray}
The magnetic field is given by ($\mu_0=$ vacuum magnetic permeability, $\epsilon=$ electric permittivity of dielectric medium)
\begin{equation} \label{H}
{\bf H}= -i\sigma \, \sqrt{\frac{\epsilon}{\mu_0}}\,{\bf E}.
\end{equation}

The expressions above for ${\bf E}$ and  ${\bf H}$ are {\it exact} solutions of the Maxwell equations and  approximate solutions of the boundary conditions corresponding to Kirchhoff's classical diffraction theory \cite{RWprelim}. This is in sharp contrast with the usual paraxial models, where approximate solutions of the Maxwell equations are employed and diffraction is completely neglected (the beam waist size is usually assumed to be much smaller than the  transverse sizes of the optical elements).  

The electric and magnetic fields are linear combinations of terms of the form $e^{i(\ell +m)\phi}I^{(\ell)}_{m}$. Since the phase factor is ill-defined on the beam axis (phase singularity), consistency requires that the functions  $I^{(\ell)}_{m}$ vanish at $\rho=0$, {\it except} when $\ell+m=0$. This important property is verified by replacing the result $ J_n(0)=\delta_{n,0}$ into (\ref{I}).

The electric and magnetic field components $E_z$ and $H_z$ have no phase singularity when $\ell=\pm 1$ and $\sigma=-\ell$ (anti-parallel incident beam orbital and spin angular momenta). In this situation, the vectors ${\bf E}$ and $\bf H$ oscillate along the beam axis in phase quadrature. 

There is a second situation with non-vanishing axial intensity: $\ell=\pm 2,$ $\sigma=\mp 1$. From (\ref{Ex}) and (\ref{Ey}), we find, on the axis, $E_y = -i\sigma E_x$ and $E_z=0$, so that the field on the axis is circularly polarized as in the entrance port. However, the sense of rotation is opposite to that at the entrance port. 

\begin{figure}[h]
\centering
\includegraphics[height=6.5cm]{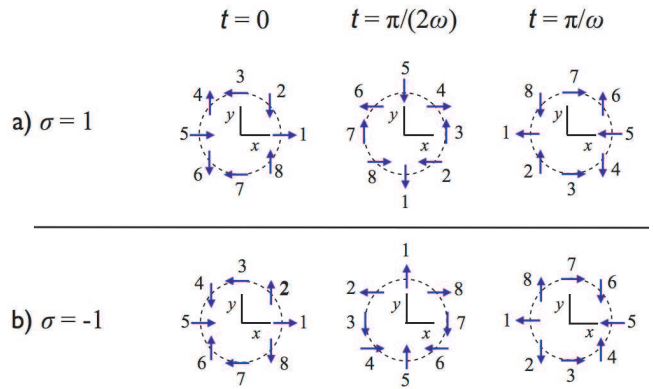}
\caption{ (color online)
Electric field orientations for the paraxial  ${\rm LG}_{02}$ beam  (before the objective), with 
(a) left-handed ($\sigma = 1$) and (b)   right-handed ($\sigma=-1$) 
circular polarizations, at times $t=0,$ $t=\pi/(2\omega)$ and $t=\pi/\omega.$ The resulting spatial patterns counter-rotate with respect 
to the circular polarization of the incident beam.}
\label{pattern}
\end{figure}

To provide geometrical insight into the origin of the reversed rotation, we analyze the effects of strong focusing
($\theta_0\gg 1$) on vector interference at the focal point. Fig. \ref{pattern}
 shows the electric field vectors for the incident paraxial ${\rm LG}_{02}$ beam on  a plane parallel to the $xy$
  plane. The objective maps each of these vectors to the focal point by three-dimensional transport, constrained by the condition that its angle with respect to the plane of incidence (containing $\bk$ and the axis) is conserved.
Thus, the cylindrical 
 azimuthal components (which are perpendicular to the incidence plane) are transported parallel to themselves, whereas the 
 cylindrical radial components yield a $z$ component proportional to $\sin{\theta}$ and a component on the $xy$
  plane proportional to $\cos{\theta}$.
  In the following qualitative discussion
  for $\theta_0\gg 1,$ we consider a large value of $\theta$, 
  so that the latter contribution is small. 
At $t = 0,$ vectors 3 and 7 in Fig. \ref{pattern} contribute $2E_{0}$ along the negative $x$ direction, whereas 1 and 5 contribute $2E_{0}\cos{\theta}\ll 
2E_{0}$ along the positive $x$-direction. The contributions to the $z$-component
 from opposite points (1 and 5, 2 and 6, etc) are exactly cancelled. The diference between 
 $\sigma = +1/-1,$ corresponding to 
 Figs. 1(a) and 1(b),  lies in the contributions of vectors 2, 4, 6 and 8. 
 When adding their azimuthal components, we find a vector 
 pointing along the positive $x$-direction in case (a), and then the field
 resulting from interference with 3 and 7
  vanishes.
 On the other hand,  they add to the contributions of 3 and 7 
 in case (b), and the resulting electric field, 
 containing a  factor $(1-\cos\theta)$ in agreement with
 (\ref{Ex}), (\ref{Ey}) and (\ref{f2}), points
 along  the negative $x$-direction at $t=0.$
 Moreover, 
  while the electric field at each spatial position on the incident beam rotates clockwise in Fig. \ref{pattern}b,
   the resulting overall pattern rotates counter-clockwise, with the same angular frequency. Thus, the resulting vector sum at the focal point also rotates counter-clockwise.
  
These remarkable effects  disappear in the paraxial focusing
 limit, which may be obtained from our more general results by assuming that $\theta_0\ll 1$ (small NA) \cite{foot_paraxial_limit}. In this limit, the angular functions given by (\ref{f0})-(\ref{f2}) satisfy $f_2(\theta)\approx \theta^2/2\ll f_1(\theta)\approx \theta \ll f_0(\theta)\approx 2$. Then, we may neglect $I^{(\ell)}_{2\sigma}$ and $I^{(\ell)}_{\sigma}$ in (\ref{Ex})-(\ref{Ez}), yielding $E_z\ll E_y\approx i\sigma E_x$: the polarization of the incident beam is preserved in this case. Moreover, the intensity along the beam axis is very small even if $\ell+\sigma=0$ or if $\ell+2\sigma=0$ (specially in the second case).

Polarization thus plays a minor role in the paraxial limit. On the other hand, there is a nontrivial interplay between polarization and the field  spatial dependence in the nonparaxial regime. To fully picture the polarization of the focused beam, it is convenient to derive the cylindrical  field components \cite{zhan} from (\ref{Ex}) and (\ref{Ey}):
\begin{eqnarray}
E_{\rho}(\rho,\phi,z) &=&  (-i)^{\ell+1} E_0 e^{i(\ell+\sigma)\phi} \left( 
 I^{(\ell)}_{0} +  I^{(\ell)}_{2\sigma}  \right)  \label{Erho} \\
E_{\phi}(\rho,\phi,z) &=&  \sigma (-i)^{\ell} E_0 e^{i(\ell+\sigma)\phi} \left( 
 I^{(\ell)}_{0} -  I^{(\ell)}_{2\sigma}  \right). \label{Ephi}
\end{eqnarray}

Thus, all cylindrical components, including $E_z$ given by (\ref{Ez}), have the same dependence on $\phi$, determined by the total axial incident beam angular momentum $\ell+\sigma$. Note, however, that the electric and magnetic  fields depend separately on the spin and orbital indexes $\sigma$ and $\ell$, since their dependence on $\rho$ and $z$ is determined by the functions $I^{(\ell)}_{m}(\rho,z)$ [Eq. (\ref{I})]. These functions obey the important symmetry relation
\begin{equation}\label{transf1}
I^{(-\ell)}_{-m}(\rho,z) = (-)^{\ell+m} I^{(\ell)}_m(\rho,z).
\end{equation}
It follows that, when we reverse {\it both} incident  spin and orbital indexes ($\ell\rightarrow -\ell, \sigma  \rightarrow -\sigma $) the electric field changes as 
\begin{eqnarray}
\label{transf2}
E_{\rho}(\rho,\phi,z) & \longrightarrow
& E_{\rho}(\rho,-\phi,z) \\
\label{transf3}
E_{\phi}(\rho,\phi,z) & \longrightarrow
& - E_{\phi}(\rho,-\phi,z) \\
\label{transf4}
E_{z}(\rho,\phi,z) & \longrightarrow
&  E_{z}(\rho,-\phi,z),
\end{eqnarray}
whereas corresponding results for the magnetic field have the opposite sign due to multiplication by $\sigma$ in Eq. (\ref{H}). These transformation rules allow us to extend to negative values of $\ell$ results from the discussion for $\ell\ge 0$, presented below. 

\begin{figure}[h]
\centering
\includegraphics[width=9cm]{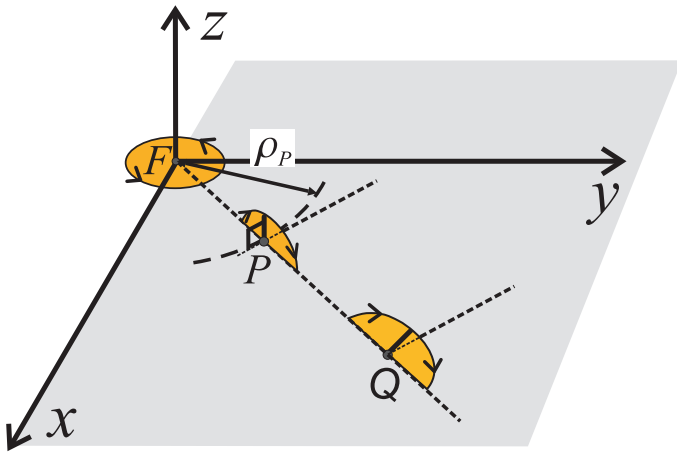}
\caption{(color online)
Polarization on the focal plane $xy$ for $\ell=2,$ $\sigma=-1$. Along the circle $\rho=\rho_P$, the polarization ellipse is perpendicular to the focal plane (as illustrated at point $P$). Inside the circle,  $\rho<\rho_P,$ the projection of the electric field onto the $xy$ plane rotates counterclockwise, so that the 
rotation is reversed with respect to the polarization of the paraxial incident beam. At the focal point $F$ itself, the polarization is circular. }
\label{fig-pola}
\end{figure}

On the focal plane $z=0,$ all functions $I^{\ell}_m$ are real according to (\ref{I}). Then, from (\ref{Ez}) and (\ref{Erho})-(\ref{Ephi}) we conclude that $E_{\phi}$ and $E_z$ are in phase or in phase opposition depending on the sign of $( I^{(\ell)}_{0} -  I^{(\ell)}_{2\sigma})/I^{(\ell)}_{\sigma} $, whereas $E_{\rho}$ is in phase quadrature. The polarization ellipse is at an angle $\alpha=\arctan[I^{(\ell)}_{\sigma}/(I^{(\ell)}_{0} -  I^{(\ell)}_{2\sigma})]$ with respect to the focal plane, with its principal axis along the $\boldsymbol{\hat \rho}$ direction. This is illustrated at point $Q$ in Fig. \ref{fig-pola}, for $\ell=2,$ $\sigma=-1.$ 

\begin{figure}[h]
\centering
\includegraphics[width=9.5cm]{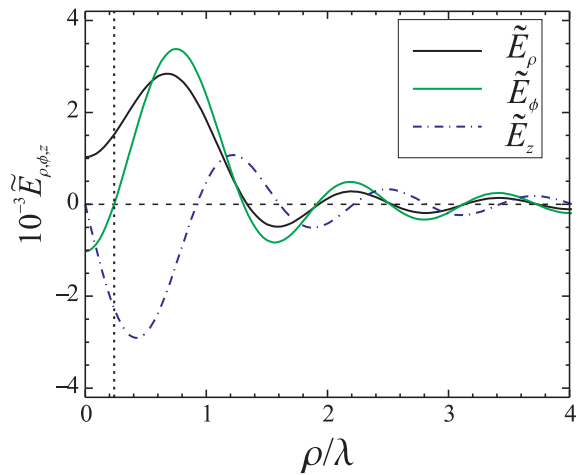}
\caption{(color online)
Electric field cylindrical components   on the focal plane divided by the maximum incident beam amplitude
 versus  $\rho/\lambda$, with $\ell=2,$ $\sigma=-1,$  NA$= 1.25$, $n=1.5$, $\gamma = 1.4$, 
 $T_{\rm obj}=0.7$, $w_0=3\,{\rm mm}$ and $\lambda=0.5\,\mu{\rm m}$.
 The vertical dotted line indicates the circle $\rho=\rho_P$.}
\label{amplitudes}
\end{figure}

For a detailed picture of the polarization in this interesting case, we plot in Fig. \ref{amplitudes}  the field amplitude 
cylindrical
components as functions of $\rho/\lambda.$ The amplitudes are divided by the maximum incident electric field amplitude at the entrance port, obtained from Eq. (\ref{port}) on the circle of maximum intensity $\rho= w_0$ \cite{PadgettAllen1995}. These dimensionless amplitudes are given by
\begin{eqnarray}
\label{Etilderho}
{\tilde E}_{\rho,\phi} &=
& \frac{\pi e}{2\sqrt{2}}\,\gamma\,T_{\rm obj}\,\frac{w_0}{\lambda}\left( I_0^{(2)} \pm I_{-2}^{(2)} \right) \\
\label{Etildephi}
{\tilde E}_{z} &=
& -\frac{\pi e}{\sqrt{2}}\,\gamma\,T_{\rm obj}\,\frac{w_0}{\lambda} \,I_{-1}^{(2)}
\end{eqnarray}
The pre-factor $w_0/\lambda$ in (\ref{Etilderho})-(\ref{Etildephi}) arises from the focused beam intensity enhancement by the geometric factor $w_0^2/\lambda^2,$ which is of order of the ratio between the beam transverse areas at the entrance port and at  the focal plane  \cite{filling_factor}, as expected by energy conservation.

\begin{figure}[h]
\centering
\includegraphics[width=8cm]{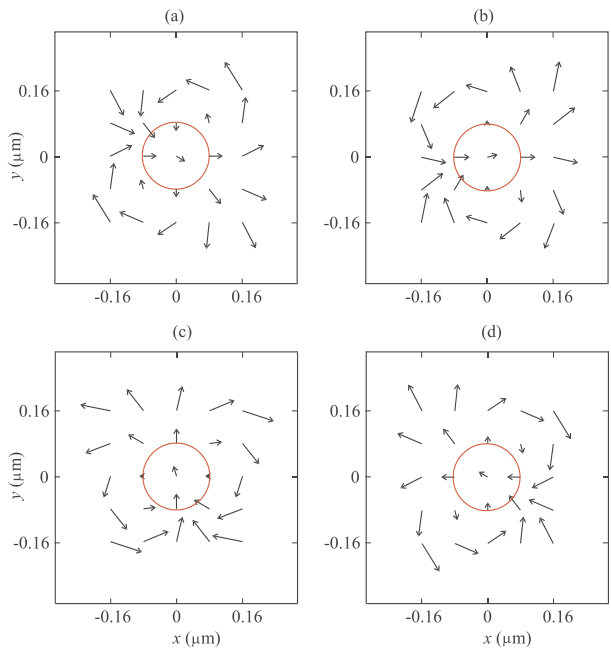}
\caption{(color online)
Electric field orientations on the focal plane at times (a) $t=\pi/(3\omega),$ (b) $t=7\pi/(12\omega)$, (c) 
$t=5\pi/(6\omega)$ and (d) $t=13\pi/(12\omega).$ We take $\ell=2,$ $\sigma=-1.$ 
The red (gray) line indicates the circle $\rho=\rho_P$.
The electric field at a given spatial position outside this circle
rotates clockwise like the incident beam. Inside the circle, however, the 
rotation is counterclockwise, as illustrated by the vector  at the focal  
point. 
}
\label{cpoeletrico}
\end{figure}

For the numerical example shown in Fig.~\ref{amplitudes}, we take NA$= 1.25$, $n=1.5$, $T_{\rm obj}=0.7$, $\gamma = 1.4$, $w_0=3\,{\rm mm}$ and $\lambda=0.5\,\mu{\rm m}$. As expected, the field is localized in a region around the focus, with dimensions of the order of the wavelength $\lambda$. When the amplitudes ${\tilde E}_{\rho}$ and ${\tilde E}_{\phi}$ have the same sign, the projection of the electric field on the $xy$ plane rotates clockwise, like the incident beam. This happens in most spatial regions, in particular around the ring of maximum intensity (to be discussed in the next section). On the circle $\rho=\rho_P=0.24\,\lambda$ (for the above numerical values), we have $E_{\phi}=0 $, so that the polarization ellipse is perpendicular to the $xy$ plane ($\alpha=\pi/2$). This is represented at point $P$ in figure \ref{fig-pola}. Inside the disk ($\rho<\rho_P$), ${\tilde E}_{\rho}$ and ${\tilde E}_{\phi}$ have opposite signs; thus, the projection of ${\bf E}$ on the $xy$ plane now rotates counter-clockwise. At the focal point ($\rho=0$), the polarization is circular with reversed rotation as already explained.
These properties are illustrated by Fig.~\ref{cpoeletrico}, showing the projection of the electric field vectors 
onto the focal plane at different times. 

The same reversion takes place for $\ell>2,$ and the radius $\rho_P$ increases linearly with $\ell$. However, there is a phase singularity on the beam axis in these cases, so that the field vanishes at 
$\rho=0$ instead of being circularly polarized. On the other hand, no reversion occurs when $\ell$ and $\sigma$ have the same sign, nor with $\ell=\pm 1$. Thus, we may interpret the reversion as a nontrivial interplay between polarization and spatial structure, 
as already discussed in  connection with Fig.~\ref{pattern}. 
Near the beam axis ($\rho< \rho_P$), for $|\ell|\ge 2$, the spatial angular dependence dominates over the polarization inherited from the incident beam, 
leading to a reversed rotation when $\ell$ and $\sigma$ have opposite signs. 

In the next section, we show that the direction of energy flow near the axis also gets {\it inverted} with respect to the overall propagation direction (defined by the paraxial incident beam) when $\ell$ and $\sigma$ have opposite signs and $|\ell|\ge 2.$

\section{Energy density and flux}

The (time-averaged) electric energy density $u_E=(\epsilon/4){\bf E}\cdot{\bf E}^*$ is symmetrical with respect to the focal plane $z=0:$ $u_E(\rho,-z)=u_E(\rho,z),$ independent of $\phi$ (rotational symmetry) and identical to the magnetic energy density  at each point: $u_E=u_M$ (these last two properties would not hold for linear polarization of the incident beam). From (\ref{Ez}), (\ref{Erho}) and (\ref{Ephi}), we find for the total energy density $u=u_E+u_M$
\begin{equation}
u(\rho,z)= \epsilon E_0^2\left(|I_0^{(\ell)}|^2+2|I_{\sigma}^{(\ell)}|^2+|I_{2\sigma}^{(\ell)}|^2\right)
\end{equation}
From (\ref{transf1}) or (\ref{transf2})-(\ref{transf4}), it is clear that the energy density is invariant under change of signs of  both $\ell$ and $\sigma.$ Thus, results for negative values of 
$\ell$ are easily obtained from those for $\ell>0$ (discussed below) by simply inverting $\sigma.$

\begin{figure}[h]
\centering
\includegraphics[width=9.5cm]{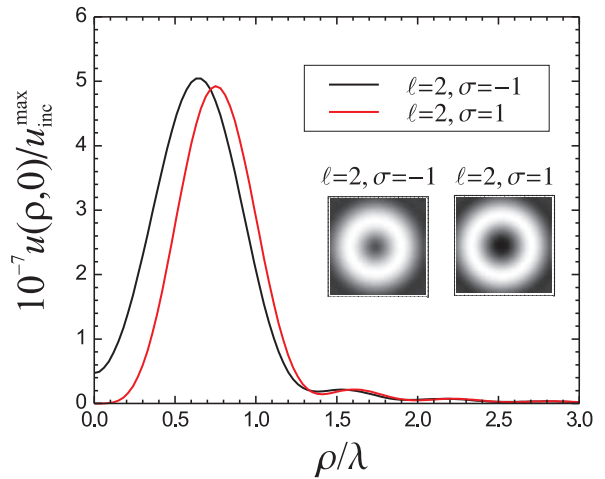}
\caption{(color online) Energy density ratio
$u(\rho,0)/u_{\rm inc}^{\rm max}$
 versus $\rho/\lambda$ at the focal plane (same parameters as in Fig.~\ref{amplitudes}). }
\label{energy}
\end{figure}

In Fig. \ref{energy}, we plot the ratio 
$u(\rho,0)/u_{\rm inc}^{\rm max}$ as a function of $\rho/\lambda$, for the same parameters as in Fig.~\ref{amplitudes} ($u_{\rm inc}^{\rm max}$ represents the maximum density of the incident paraxial beam). We also show, for comparison, corresponding results for $\ell=2$, $\sigma=1$, with vanishing density at the axis. Although they have the same spatial profile at the objective entrance port, the strong interplay with polarization results in different distributions on the focal plane. For $\ell=2$, $\sigma=-1$, the bright ring is thicker  and slightly displaced inwards (see also the insets in Fig.~\ref{energy}).

\begin{figure}[h]
\centering
\includegraphics[width=9.5cm]{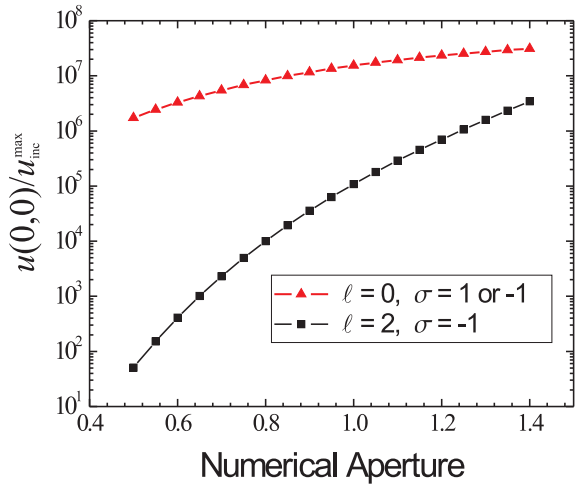}
\caption{(color online) 
Energy density ratio
$u(0,0)/u_{\rm inc}^{\rm max}$ at the focal point versus NA. Same parameters as in Fig.~\ref{amplitudes}.}
\label{NA}
\end{figure}

 The density value at the focus 
 for $\sigma=-1$
 is only $9\%$ of its maximum on the circle $\rho\approx 0.65\,\lambda$, but it is  very large in comparison with the incident beam energy density:
  $u(0,0)/u_{\rm inc}^{\rm max}=4.7\times 10^6.$ This arises partly from focusing the available energy to a much tighter region, with an area of the order of $\lambda^2$ for the large NA objective considered in this example. But increasing NA has a stronger effect on the focal energy density for this particular case (and also for $\ell+\sigma=0$). This is shown in Fig.~\ref{NA}, where we plot the ratio $u(0,0)/u_{\rm inc}^{\rm max}$ as a function of NA for 
$\ell+2\sigma=0$. We also show the results for $\ell=0$: in this case the effect of increasing NA is less dramatic because the axial density does not vanish in the paraxial limit. 

\begin{figure}[h]
\centering
\includegraphics[width=8cm]{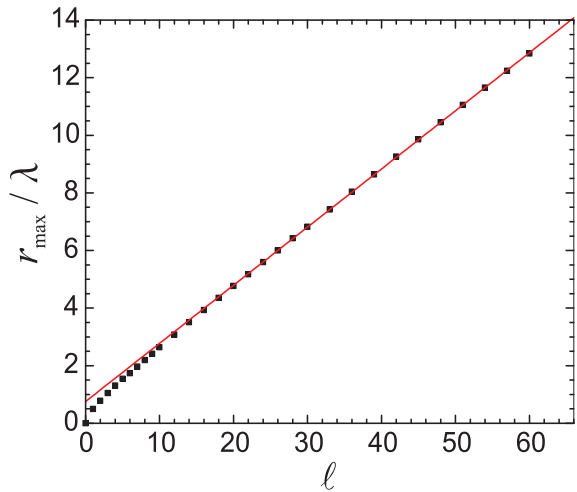}
\caption{(color online) Radius of maximum energy density (divided by the wavelength) versus orbital angular momentum index $\ell$, with $\sigma=1$. The result of a linear fit for $\ell>20$ is also shown ($r_{\rm max}/\lambda=0.202\,\ell+0.758$). }
\label{rmax}
\end{figure}

For all $\ell\neq 0,$ the peak value of  the energy density on the focal plane is reached at a radius that increases with $\ell.$ For a paraxial Laguerre-Gaussian beam, this radius grows as $\sqrt{\ell}$ (for given values of $w_0$ and $\lambda$). Fig.~\ref{energy} shows that the radius $r_{max}$ of a tightly focused  beam also depends on polarization. Remarkably, for a given polarization it grows 
 linearly with $\ell$ for large $\ell$, as shown experimentally by Curtis and Grier  from the rotational dynamics of optically trapped particles~\cite{GrierPRL2003}. In Fig. \ref{rmax}, we show that the beam radius is in fact well fitted by a linear function for $\ell>20.$ 

\begin{figure}[h]
\centering
\includegraphics[width=7.5cm]{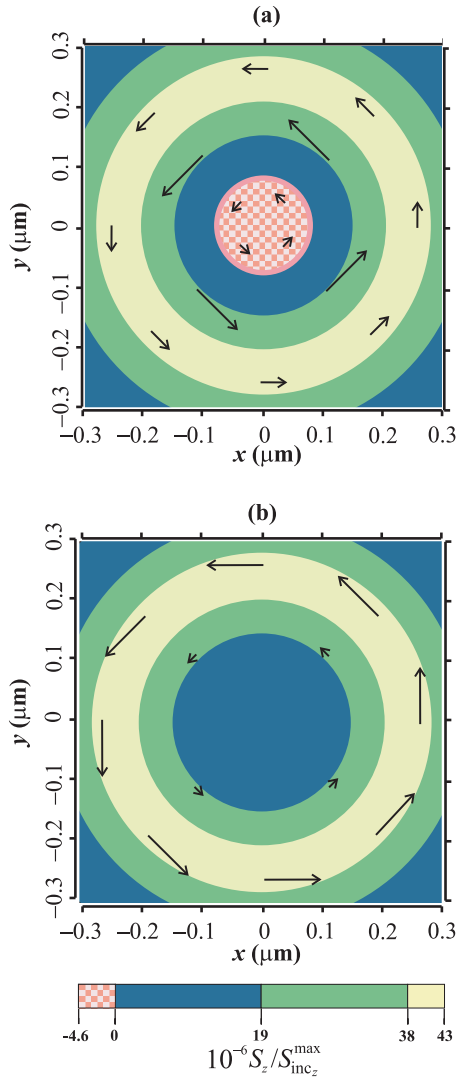}
\caption{(color online) 
The arrows represent the projection of the Poynting vector on the focal plane, for
 $\ell=2$ and (a) $\sigma=-1$, (b) $\sigma=1$
(same parameters as in Fig.~\ref{amplitudes}).
The values of $S_z/S_{{\rm inc}z}^{\rm max}$ are represented by the false color map 
($S_{{\rm inc}z}^{\rm max}=$ maximum Poynting vector  of the incident beam).
In case (a), $S_z$ is negative within the circle 
 $ \rho = \rho_P$ .  
 }
\label{S}
\end{figure}

The Poynting vector
\begin{equation}
{\bf S}({\bf r})= \frac{1}{2}\, {\rm Re}\left( {\bf E}\times {\bf H}^* \right)=
 -\frac{\sigma}{2}\,\sqrt{\frac{\epsilon}{\mu_0}}
\,{\rm Im}\left({\bf E}\times {\bf E}^*\right)
\end{equation}
yields the energy flux density at position $\bf r$. From (\ref{Ez}) and (\ref{Erho})-(\ref{Ephi}) we find
\begin{eqnarray}
\label{Srho}
S_{\rho} & =& 2\sigma\sqrt{\frac{\epsilon}{\mu_0}}\,E_0^2\;{\rm Im}\left[
\left( 
 I^{(\ell)}_{0} -  I^{(\ell)}_{2\sigma}  \right) I^{(\ell)}_{\sigma}{}^*
\right] \\
\label{Sphi}
S_{\phi} & =&  2\sqrt{\frac{\epsilon}{\mu_0}}\,E_0^2\;{\rm Re}\left[
\left( 
 I^{(\ell)}_{0} +  I^{(\ell)}_{2\sigma}  \right) I^{(\ell)}_{\sigma}{}^*
\right] \\
\label{Sz}
S_{z} & =&   \sqrt{\frac{\epsilon}{\mu_0}}\,E_0^2\,\left(
|I^{(\ell)}_{0}|^2 -  |I^{(\ell)}_{2\sigma}|^2  \right) 
\end{eqnarray}
As expected by symmetry, the cylindrical components do not depend on $\phi.$ Since the functions 
$I^{(\ell)}_{2\sigma}$ are real at the focal plane $z=0,$ we get $S_{\rho}=0$ on this plane. This is also expected, since $S_{\rho}$ must change sign as the beam converges and then diverges from the focal plane. The polarization ellipses represented in Fig.~\ref{fig-pola} already provide the directions of the 
Poynting vector field at the focal plane: on each point, $\bf S$ is perpendicular to the corresponding ellipse. When $\ell$ and $\sigma$ have opposite signs and $|\ell|\ge 2$, Eq.~(\ref{Sz}) yields $S_z=0$  at $\rho=\rho_P$ (see Fig.~\ref{fig-pola}), so that  the Poynting vector is parallel to $\boldsymbol{\hat{\phi}}$ on this circle  (or anti-parallel 
if $\ell+\sigma<0$). Moreover, $S_z$ is {\it negative} inside the disk $\rho<\rho_P,$ so that the local energy current density flux near the axis is antiparallel to the incident beam direction. Note that the sense of the electric field rotation with respect to the local energy flux direction is everywhere the same, because within the disk $\rho<\rho_P$ both are reversed. 
Three-dimensional insight into the beam energy flow is provided by Fig. \ref{S}. The projection of the Poynting vector on the focal plane is represented by arrows, while its axial component is represented in magnitude and sign by the false color map, for
$\ell=2, \sigma=-1$ (a) and $\ell=2, \sigma=+1$ (b).
 The axial sign reversal within the circle 
 $
 \rho = \rho_P$  is apparent.

The global energy flux is of course directed along the positive $z$ axis. Using (\ref{I}), (\ref{Sz}) and the result
\begin{equation}
\label{JJ}
\int_0^{\infty} d\rho \,\rho\, J_{n}(k\rho\sin\theta)\,J_{n}(k\rho\sin\theta')=\frac{1}{k^2\sin\theta}
\,\delta(\sin\theta-\sin\theta')
\end{equation}
we find that the global energy flux across a plane corresponding to some fixed value of $z$ is independent of $z$ (as expected by energy conservation) and given by 
\begin{equation} \label{energy_flux_focal}
{\cal F} = 8\pi
\sqrt{\frac{\epsilon}{\mu_0}}
\frac{E_0^2}{k^2}
(\sqrt{2}\gamma)^{2|\ell|}
 \int_0^{\theta_0} d\theta\cos\theta(\sin\theta)^{2|\ell|+1} 
e^{-2\gamma^2\sin^2\theta}
\end{equation}
We may check this result against the incident energy flux at the objective entrance port (radius $r_{\rm obj}=f\sin\theta_0$):
\begin{equation}
\label{Fp}
{\cal F}_{\rm inc} =
2\pi \int_0^{r_{\rm obj}} d\rho\, \rho \, \mathbf{\hat z}\cdot{\bf S}_{\rm inc}
\end{equation}
From (\ref{port}), (\ref{E0}) and (\ref{energy_flux_focal}), we find ${\cal F}= nT_{\rm obj}^2{\cal F}_{\rm inc}$
as expected. As a final remark about the Poynting vector, we note the transformation rules under inversion of both spin and orbital indices 
($\ell\rightarrow -\ell, \sigma  \rightarrow -\sigma $) obtained from (\ref{transf1}) and (\ref{Srho})-(\ref{Sz}). We find that $ S_{\rho} $  and $ S_{z} $ are invariant whereas 
\begin{equation}
\label{tSphi}
S_{\phi}  \longrightarrow
 - S_{\phi}.
\end{equation}
This sign change is expected  since $S_{\phi}$ is  directly related to the optical angular momentum as discussed in the next section.  

\section{Angular momentum}

\begin{figure}[h]
\centering
\includegraphics[width=7cm]{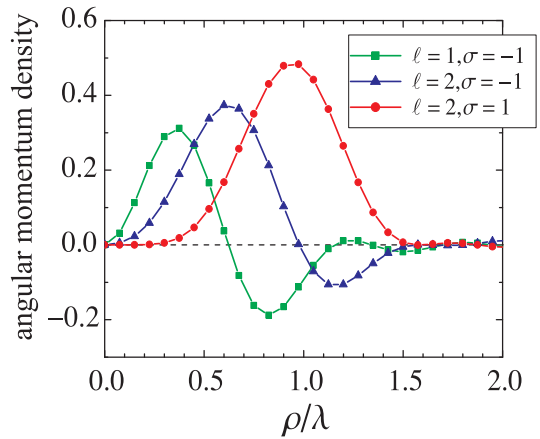}
\caption{ (color online) Dimensionless pseudo-angular momentum density ${\tilde j}_z$ 
 on the focal plane
versus $\rho/\lambda$ (same parameters as in Fig.~\ref{amplitudes}).}
\label{angularmomentum}
\end{figure}

From the Minkowski linear pseudo-momentum density \cite{Brevik} ${\bf g} = {\bf D}\times{\bf B},$ we get the pseudo-angular momentum density \cite{KristensenWoerdman} 
\begin{equation}
\label{jz}
j_z(\rho,z)= n^2 \rho S_{\phi}/c^2, 
\end{equation}
which may be numerically calculated from (\ref{I}) and (\ref{Sphi}). In Fig.~\ref{angularmomentum}, we plot the dimensionless density ($u_{\rm max}=$ energy density at the focus for $\ell=0$) 
\[
{\tilde j}_z = \frac{\omega\, j_z}{u_{\rm max}}
\]
as a function of $\rho/\lambda$ for several values of $\ell$ and $\sigma.$ According to Eq.~(\ref{tSphi}), negative values of $\ell$ for a given polarization $\sigma$ are be obtained simply by changing the sign of ${\tilde j}_z$ calculated for $-\ell, -\sigma.$

The results shown in Fig.~\ref{angularmomentum} are qualitatively similar to those for a paraxial beam, with $\lambda$ playing the role of the paraxial beam waist $w_0$ as a transverse length scale for our diffraction-limited beam. For a paraxial beam, the orbital contribution is proportional to $\ell\,|E|^2$, whereas the spin  is proportional to $-\sigma\rho\partial_{\rho}|E|^2/2$ \cite{Allen92}. For $\sigma=-1,$ Fig.~\ref{angularmomentum} shows that the pseudo-angular momentum density is in fact negative near the outer edge of the ring of maximum intensity ($\partial_{\rho}|E|^2<0$) as in the paraxial case. Closer to the beam axis, on the other hand, the density is always positive for  $\ell>0$, so that one might be tempted to conclude that the orbital contribution dominates near the beam axis. 

However,  the identification of spin and orbital contributions  is not straightforward (see Refs.~\cite{Cohen} and \cite{VanEnk} for a general discussion), particularly in the nonparaxial regime, as shown by 
Barnett and Allen \cite{Barnett94}. They took a general nonparaxial cylindrically symmetric beam 
containing a phase factor $e^{i\ell\phi}$. Besides the ambiguity in the identification of spin and orbital contributions,  the ratio between the total angular momentum per unit length and the total energy per unit length was found  not to be $(\ell+\sigma)/\omega$ as in the case of a paraxial beam. 

The nonparaxial focused beam model, as derived here, does not coincide with the Barnett-Allen model. The reason is that the angular position phase factor is not simply $e^{i\ell\phi}:$ the Cartesian electric field components contain instead all three phase factors $e^{i\ell\phi}$, $e^{i(\ell+\sigma)\phi}$ and $e^{i(\ell+2\sigma)\phi}$ according to (\ref{Ex})-(\ref{Ez}). Nevertheless, we find the same ambiguity concerning the separation between spin and orbital contributions to the pseudo-angular momentum density in the dielectric medium, and no simple relation between pseudo-angular momentum and energy per unit length.

A more convenient description of optical angular momentum flow outside the paraxial regime is provided by the angular momentum current flux density, represented by a second order tensor $M_{ij}$~\cite{Jackson}. With the pseudo-angular momentum density given by (\ref{jz}) in terms of the Minkowski linear pseudo-momentum \cite{KristensenWoerdman}, the flux density of the axial pseudo-angular momentum component along the $z$ direction is
\begin{equation}\label{Mzz}
M_{zz} = \frac{1}{2} \, {\rm Re} \Bigl[
y\left(\epsilon E_x E_z^* +\mu_0 H_x H_z^* \right) - x
\left(\epsilon E_y E_z^* +\mu_0 H_y H_z^* \right)
    \Bigr].
\end{equation}
It is useful to write $E_z$ and $H_z$ in (\ref{Mzz}) in terms of the curl of $\bf H$ and $\bf E$  by using the Maxwell equations. The factor $\epsilon$ appearing in (\ref{Mzz}) is canceled, and the resulting expression is further simplified by using Eq.~(\ref{H}) to write the magnetic field in terms of the electric field. 

Barnett has shown that the total angular momentum flux 
\begin{equation}
{\cal M} = \int_0^{\infty}d\rho\rho
\int_0^{2\pi}d\phi  \,M_{zz}
\end{equation}
through a plane parallel to the $xy$ plane can be separated into two gauge-independent contributions respectively associated to the polarization and to the spatial field structure \cite{Barnett02}. These two contributions provide natural definitions for spin and orbital angular momentum fluxes, 
respectively. After integration by parts, one finds: 
\begin{equation}
{\cal M}  =  {\cal M}^{\rm spin}+{\cal M}^{\rm orb}  
\end{equation}
\begin{equation}
{\cal M}^{\rm spin}  =     \frac{\sigma}{2\omega}\sqrt{\frac{\epsilon}{\mu_0}} \int_0^{\infty}d\rho\rho
\int_0^{2\pi}d\phi\left(|E_x|^2+|E_y|^2\right) \label{Mspinpre}
\end{equation}
\begin{eqnarray}
{\cal M}^{\rm orb} &=&   \frac{\sigma}{2\omega}\sqrt{\frac{\epsilon}{\mu_0}} \int_0^{\infty}d\rho\rho
\int_0^{2\pi}d\phi \,{\rm Re}\bigl(E_y\partial_{\phi}E_x^*  \nonumber \\
 & &   -E_x\partial_{\phi}E_y^*\bigr)  \label{Morbpre}
\end{eqnarray}
Since the component $E_y$ already contains a factor $\sigma$ according to (\ref{port}) and (\ref{Ey}), 
the pre-factor $\sigma$ in the expression on the r.-h.-s. of (\ref{Morbpre}) is canceled as expected. 

If one applies (\ref{Mspinpre}) and (\ref{Morbpre}) to a paraxial LG beam, one finds, after division by the energy flux
 ${\cal F}_{\rm inc}$, ${\cal M}^{\rm spin}_{\rm inc}/{\cal F}_{\rm inc}= \sigma/\omega$ and ${\cal M}^{\rm orb}_{\rm inc}/{\cal F}_{\rm inc}= \ell/\omega.$ Thus, the pseudo-angular momentum  per photon in the dielectric medium does not depend on $\epsilon$ and is exactly the same as in vacuum, as already verified experimentally \cite{KristensenWoerdman}.  

We now apply (\ref{Mspinpre}) and (\ref{Morbpre}) to the nonparaxial focused beam. By writing the electric field components in terms of $I_m^{(\ell)}(\rho,z)$ as in (\ref{Ex}) and (\ref{Ey}), we find 
\begin{eqnarray}
\label{Mspin}
{\cal M}^{\rm spin} & =    \frac{2\pi\sigma}{\omega}\sqrt{\frac{\epsilon}{\mu_0}}E_0^2 
\int_0^{\infty}d\rho\rho
&\left(|I_0^{(\ell)}|^2+|I_{2\sigma}^{(\ell)}|^2\right) \\
{\cal M}^{\rm orb} & =   \frac{2\pi}{\omega}\sqrt{\frac{\epsilon}{\mu_0}} E_0^2  \int_0^{\infty}d\rho\rho
 &\,\Bigl[\ell\left( |I_0^{(\ell)}|^2-|I_{2\sigma}^{(\ell)}|^2\right)  \nonumber\\
\label{Morb}
 &   &-2\sigma |I_{2\sigma}^{(\ell)}|^2\Bigr]
\end{eqnarray}
Using the definition (\ref{I}) and the result (\ref{JJ}), it is straightforward to show that both ${\cal M}^{\rm spin}$ and ${\cal M}^{\rm orb}$ are independent of $z,$ so that the spin and orbital fluxes are independently conserved as the focused beam propagates in the region beyond the objective. However, their values do not separately coincide with the spin and orbital fluxes per unit power for the incident paraxial beam before the objective. Because of the angular dependence $e^{i(\ell+2\sigma)\phi}$ of both $E_x$ and $E_y$ [cf. (\ref{Ex}) and (\ref{Ey})], the angular derivative contained in the orbital flux in (\ref{Morbpre}) leads to the term proportional to $2\sigma$ in (\ref{Morb}). 

Only the total pseudo-angular momentum flux per unit power is conserved by focusing, 
as expected given the cylindrical symmetry of the objective. In fact, when adding the spin and orbital fluxes given by (\ref{Mspin}) and (\ref{Morb}), we find, by comparison with Eq.~(\ref{Sz}), 
\begin{equation}
{\cal M}^{\rm orb}+{\cal M}^{\rm spin}=\frac{(\ell+\sigma)}{\omega}\,{\cal F}.
\end{equation}

The relative change of the spin flux per unit power with respect to the incident beam is 
\begin{equation}
r=\frac{{\cal M}^{\rm spin}/{\cal F}-{\cal M}^{\rm spin}_{\rm inc}/{\cal F}_{\rm inc}}
{{\cal M}^{\rm spin}_{\rm inc}/{\cal F}_{\rm inc}},
\end{equation}
whereas the relative change of orbital flux per unit power is simply $-(\sigma/\ell)r.$  From (\ref{JJ}), (\ref{Mspin}) and (\ref{Morb}), we find
\begin{equation}
\label{r}
r=\frac{2\int_0^{\theta_0} d\theta\, g_{|\ell|}(\theta) \sin^4(\theta/2)}
{\int_0^{\theta_0}  d\theta\, g_{|\ell|}(\theta) \cos\theta} 
\end{equation}
where
\begin{equation}
\label{g}
g_{n}(\theta)  \equiv \exp(-2\gamma^2\sin^2\theta)\left(\sin\theta\right)^{2n+1}
\end{equation}

By inspection of (\ref{r}) and (\ref{g}), we conclude that $r$ is positive and does not depend on the sign of $\ell$ nor on the value of $\sigma.$  Hence the variation of ${\cal M}^{\rm spin}/{\cal F}$ has the sign of $\sigma$ and is such that its absolute value always increases. At first sight, values for ${\cal M}^{\rm spin}/{\cal F}$ outside the range from $-1/\omega$ to $1/\omega$ might seem to contradict the fact that the photon has spin one. Note, however, that the quantities ${\cal M}^{\rm spin}$ and ${\cal F}$ represent  overall net flux balances of the local spin angular momentum and energy  across a plane of constant $z,$ in a situation where the polarization and the Poynting vector vary strongly from point to point.  As an illustration, consider the case $\ell=2,$ $\sigma=-1$ discussed at length in Sec. II.  As shown by Fig.~1, the electric field rotates counterclockwise (as seen from $z>0$) near the axis, and, as discussed in Sec. III, the energy locally flows along the negative $z$-direction. A picture in terms of ray contributions
 then goes as follows: the  flux of rays  $R_{\rm far}$  crossing the plane far from the axis contributes $R_{\rm far} \hbar \omega$ to ${\cal F}$ and $ -R_{\rm far} \hbar $ to ${\cal M}^{\rm spin}.$ On the other hand,  rays crossing near the axis (flux $R_{\rm near}<R_{\rm far}$) provide a negative contribution $-R_{\rm near} \hbar \omega$ to $\cal F$ as well as a {\it negative} contribution $- R_{\rm near} \hbar $  to ${\cal M}^{\rm spin}$ because they propagate along the negative $z$ direction with a positive $z$ component of spin angular momentum (the crucial point here is that the angular momentum is a vector quantity whereas the energy is a scalar). The net overall ratio is then
\[
\frac{{\cal M}^{\rm spin}}{\cal F} = - \frac{R_{\rm far}+R_{\rm near}}{R_{\rm far}-R_{\rm near}}\,\frac{1}{\omega}<-\frac{1}{\omega}
\]

The interplay between spin and orbital fluxes is clearly a nonparaxial effect which becomes enhanced as $|\ell|$ increases, because larger values of $|\ell|$ reduce the contribution of paraxial angles $\theta\ll 1$ in (\ref{r}). The integrands in (\ref{r}) are equal at $\theta_m\approx 74.5^o$, which corresponds to a numerical aperture NA$=1.44$ for $n=1.5.$ Typical high-NA values  $1.25$ and $1.4$ are such that for all values of $\theta$ in (\ref{r}) the integrand in the numerator is considerably smaller than that in the denominator, leading to small values for the relative change $r$. In Table I, we show the values of $r$ for NA$=1.25$ and $1.4$, for different values of $\ell.$ The orbital flux per unit power 
\begin{equation}
{\cal M}^{\rm orb}/{\cal F}=-(\sigma/\omega)\,r + (\ell/\omega)
\end{equation}
increases (decreases) when $\sigma=-1(+1).$ This seems to be in contradiction with the concept of `conversion' between spin and orbital angular momenta \cite{Zhao2007}. Note, however, that the connection between the global angular momentum flux calculated here and the mechanical effects on local probe particles is not straightforward, particularly when local quantities vary strongly, as discussed in the previous paragraph. To our knowledge, a consistent identification of spin and orbital contributions to the {\it local} angular momentum  or flux densities of nonparaxial beams is not available \cite{Barnett94} \cite{Barnett02} (this is the reason why we employ the term `interplay' rather than `exchange' or `interconversion'). 

\begin{table}
\caption{\label{table}Relative spin flux change $r$ for different values of $\ell.$ 
Additional parameters as in Fig.~3.
}
\bigskip\bigskip
\bigskip

\begin{tabular}{ccc}
\hline\hline
$\ell$ &\;\;\;\;\, NA$=1.25$\;\;\;\;\, & NA$=1.4$ \\
\hline
0&  $0.0165$ & $0.0264$ \\
1&  $0.0382$ & $0.0703$ \\
2&  $0.0579$ & $0.118$ \\
3&  $0.0742$ & $0.162$ \\
\hline\hline
\end{tabular}
\end{table}

\section{Conclusion}

The diffraction-limited optical beam that results from focusing a circularly-polarized LG  beam by a high-NA objective has some remarkable properties, particularly when $\ell$ and $\sigma$ have opposite signs. In this situation, the interplay between polarization and angular spatial dependence leads to a strong modification of  the local polarization and of the energy propagation direction in the region near the beam axis: the energy locally propagates along the negative $z$-direction near the focal point for any $|\ell|\ge 2.$ Also, the rotation of the electric field projection onto the focal plane is reversed; e.g., when $\ell+2\sigma=0,$ the polarization at the focal point is circular with the reversed  rotation. All these effects disappear in the paraxial limit, which can be obtained from our results by taking low values of NA or a small waist size at the entrance port, thereby underfilling the objective aperture ($\gamma\gg 1.$). 

The total angular momentum flux per unit power across a plane perpendicular to the $z$-axis is conserved by the focusing effect, but not the separate spin and orbital contributions. On the other hand, they are separately conserved as the beam freely propagates beyond the objective. We have presented quantitative results for the  modification of the spin and orbital fluxes. Even for the highest possible values of NA, the relative spin flux  modification is typically small, but it grows with $\ell.$ 

Interplay between light orbital and spin angular momentum is an unfamiliar concept. Geometrical insight concerning this effect was provided in connection 
with  Fig. \ref{pattern} (cf. \cite{Bokor2005}): in the paraxial case,
 spin has to do with the  change in rotational orientation of the electric field vector at a given point,
 whereas orbital angular momentum is  associated to its change
 around a circle at a given time. 
 When focusing a circularly-polarized LG beam, the interesting patterns emerging from the combination of these two concepts 
 give rise to 
 surprising results near the beam axis of the nonparaxial beam. 
 From a physical point of view, this may be regarded as a vectorial interference effect among plane waves in the angular spectrum produced by diffraction at the objective. An analogy is provided by a (TE) or (TM) mode in a parallel-plate waveguide: it may be described in terms of multiple reflections of ordinary (TEM) plane waves against the walls (boundary effects), but vectorial interference leads to the appearance of longitudinal field components. 

The results presented in this paper have possible applications to optical trapping.  
A calculation of the optical force and torque on dielectric spheres along the lines presented here is currently under way.

\section{Acknowledgments}
We thank Nathan Viana for useful discussions.
This work was supported by the Brazilian agencies CNPq, FAPERJ, FUJB and Institutos do Mil\^{e}nio de Nanoci\^{e}ncias e de Informa\c{c}\~ao Qu\^{a}ntica.


\

\begin{thebibliography}{xxxx}

\bibitem{Ashkin86} A. Ashkin, J. M. Dziedzic, J. E. Bjorkholm and S. Chu, Opt. Lett. {\bf 11}, 288 (1986).

\bibitem{Chu86} S. Chu, J. E. Bjorkholm, A. Ashkin and A. Cable, Phys. Rev. Lett. {\bf 57}, 314 (1986).

\bibitem{Ashkin_livro} A. Ashkin, {\it Optical Trapping and Manipulation of Neutral Particles Using Lasers}, (World Scientific, Singapore, 2006). 

\bibitem{review_biology_applications} A. D. Mehta, M. Rief, J. A. Spudich, D. A. Smith, and R. M. Simmons, Science {\bf 283}, 1689  (1999). 

\bibitem{Grangier} J. Beugnon {\it et al}, Nature Physics {\bf 3}, 696 (2007).

\bibitem{RWprelim}  E. Wolf,    Proc. R. Soc. London, Ser.  A {\bf 253}, 349 (1959).

\bibitem{RW} B. Richards  and E. Wolf, Proc. R. Soc. Lond. A {\bf 253}, 358 (1959). 

\bibitem{Dorn2003} R. Dorn, S. Quabis and G. Leuchs, J. Modern Opt. {\bf 50}, 1917 (2003). 

\bibitem{tweezers_th} P. A. Maia Neto and H. M. Nussenzveig, Europhys. Lett. {\bf 50}, 702 (2000); A. Mazolli, P. A. Maia Neto, and H. M. Nussenzveig, Proc. R. Soc. London, Ser. A {\bf 459}, 3021 (2003); R. S. Dutra, N. B. Viana, P. A. Maia Neto and H. M. Nussenzveig, J. Opt. A: Pure Appl. Opt. {\bf 9}, S221 (2007).

\bibitem{tweezers_exp} N. B. Viana, M. S. Rocha, O. N. Mesquita, A. Mazolli, P. A. Maia Neto, and H. M. Nussenzveig, Appl. Phys. Lett. {\bf 88}, 131110 (2006); Phys. Rev. E {\bf 75}, 021914 (2007).  

\bibitem{Grier_Nature} D. G. Grier  {\it Nature} {\bf 424} 810 (2003).

\bibitem{Baranova83} N. B. Baranova {\it et al}, J. Opt. Soc. Am. {\bf 73}, 525 (1983); V. Y. Bazhenov, M. V. Vasnetsov and M. S. Soskin, JETP Lett. {\bf 52}, 429 (1990). 

\bibitem{Ashkin} A. Ashkin, Biophys. J. {\bf 61}, 569 (1992). 

\bibitem{Allen92} L. Allen, M. W Beijersbergen, R. J. C. Spreeuw and J. P. Woerdman, Phys Rev. A {\bf 45}, 8185 (1992). 

\bibitem{livro} L. Allen, S. M. Barnett and M. J. Padgett, eds., {\it Optical Angular Momentum} (Institute of Physics, Bristol, 2003). 

\bibitem{Bekshaev} A. Bekshaev, M. Soskin and M. Vasnetsov, arXiv:0801.2309v1 (2008). 

\bibitem{Friese96} H. He, M. E. J. Friese, N. R. Heckenberg and H. Rubinsztein-Dunlop, Phys. Rev. Lett. {\bf 75}, 826 (1995).

\bibitem{Friese_and_Simpson} M. E. J. Friese, J. Enger,  H. Rubinsztein-Dunlop and N. R. Heckenberg, 
Phys. Rev. A {\bf 54}, 1593 (1996); N. B. Simpson, K. Dholakia, L. Allen and M. J. Padgett, Opt. Lett. {\bf 22}, 52 (1997).

\bibitem{ONeil} A. T. O'Neil, A. MacVicar, L. Allen and M. J. Padgett, Phys. Rev. Lett. {\bf 88}, 053601 (2002). 

\bibitem{GrierPRL2003} J. E. Curtis and D. G. Grier, Phys. Rev. Lett. {\bf 90}, 133901 (2003). 

\bibitem{Gu2003} D. Ganic, X. Gan and M. Gu, Optics Express {\bf 11}, 2747 (2003). 


\bibitem{Bokor2005}  N. Bokor, Y. Iketaki, T. Watanabe and M. Fujii, Optics Express {\bf 13}, 10440 (2005).

\bibitem{Iketaki2007}Y. Iketaki, T. Watanabe, N. Bokor and M. Fujii, Optics Lett. {\bf 32}, 2357 (2007). 

\bibitem{Cohen} C. Cohen-Tannoudji, J. Dupont-Roc and G. Grynberg, {\it Photons et atomes}
(InterEditions, Paris, 1987), ch. I. 

\bibitem{VanEnk} S. J. van Enk and G. Nienhuis, Europhys. Lett. {\bf 25}, 497 (1994). 

\bibitem{Berry} M. V. Berry, {\it Singular Optics} (eds. M. S. Soskin and M. V. Vasnetsov) SPIE {\bf 3487}, 6 (1998). 

\bibitem{Barnett94} S. M. Barnett and L. Allen, Opt. Commun. {\bf 110}, 670 (1994). 

\bibitem{Barnett02} S. M. Barnett, J. Opt. B: Quantum Semiclass. Opt. {\bf 4}, S7 (2002).

\bibitem{Zhao2007} Y. Zhao, J. S. Edgar, G. D. M. Jeffries, D. McGloin and D. T. Chiu, Phys. Rev. Lett. {\bf 99}, 073901 (2007). 

\bibitem{transmittance} N. B. Viana, M. S. Rocha, O. N. Mesquita, A.  Mazolli  and P. A. Maia Neto,  Appl. Opt. {\bf 45}, 4263  (2006).

\bibitem{foot_unitvectors} For a linearly polarized beam at the objective entrance, the polarization of each plane wave component follows from the condition that the angle between the  polarization  direction  and the incidence plane at each interface in the objective region is conserved \cite{RW}. The result for circular polarization follows by linearity. 

\bibitem{Abramowicz} M. Abramowitz and I. Stegun, {\it Handbook of Mathematical Functions} 
(Dover, New York, 1972).

\bibitem{foot_paraxial_limit} In the case of overfilling ($\gamma \ll 1$), the paraxial or low-NA limit for $\ell=0$ reproduces the standard Airy diffraction pattern of a circular aperture, with no relevant vectorial effect \cite{RW}.

\bibitem{zhan} Q. Zhan, Opt. Lett. {\bf 31}, 867 (2006).

\bibitem{PadgettAllen1995} M. J. Padgett and L. Allen, Opt. Commun. {\bf 121}, 36 (1995). 

\bibitem{filling_factor} The intensity enhancement also depends on  what fraction of the objective entrance port is filled by the incident beam. The corresponding `filling factor' is controlled by the parameter $\gamma.$
 
\bibitem{Brevik} I. Brevik, Phys. Rep. {\bf 52}, 133 (1979). 

\bibitem{KristensenWoerdman} M. Kristensen and J. P.  Woerdman, Phys. Rev. Lett. {\bf 72}, 2171 (1994). 

\bibitem{Jackson} J. D. Jackson, {\it Classical electrodynamics} (Wiley, New York, 1962). 

\end{thebibliography}
\end{document}